\documentclass[twocolumn,floatfix]{aastex61}
\usepackage{graphicx}
\usepackage[normalem]{ulem}
\usepackage{float}
\usepackage{psfig}
\usepackage[T1]{fontenc}
\usepackage[T1]{fontenc}
\usepackage{ae,aecompl}

\newcommand\aastex{AAS\TeX}

\newcommand{\radmm}{$\mathrm{rad\,m^{-2}}$}

\received{XXXX}
\revised{YYYY}
\accepted{ZZZZ}

\shorttitle{\aastex\ Regular Magnetic field inside the SQ}
\shortauthors{B. Nikiel-Wroczy\'nski et al.}

\begin{document} 

\title{A large-scale, regular intergalactic magnetic field associated with Stephan's Quintet?}

\correspondingauthor{B\l a\.zej Nikiel-Wroczy\'nski}
\email{blazej.nikiel$\_$wroczynski@uj.edu.pl}
\author{B\l a\.zej Nikiel-Wroczy\'nski}
\affil{Astronomical Observatory of the Jagiellonian University \\
ul. Orla 171, 30-244 Krak\'ow, Poland}
\author{Marian Soida}
\affil{Astronomical Observatory of the Jagiellonian University \\
ul. Orla 171, 30-244 Krak\'ow, Poland}
\author{George Heald}
\affil{CSIRO Astronomy and Space Science \\
PO Box 1130, Bentley, WA 6102, Australia}
\author{Marek Urbanik}
\affil{Astronomical Observatory of the Jagiellonian University \\
ul. Orla 171, 30-244 Krak\'ow, Poland}

\begin{abstract}

 Regular magnetic fields are frequently found within and in the outskirts of galaxies, but their presence, properties, and origin has not yet been established for galaxy groups. On the basis of broadband radio polarimetric imaging with the Westerbork Synthesis Radio Telescope (WSRT), we made use of Rotation Measure Synthesis to disentangle contributions from magnetic fields on various scales for several polarised radio sources inside, behind, or in the vicinity of the Stephan's Quintet (HCG92, SQ). We recognise the signature of a large-scale, genuinely regular, magnetised screen, seemingly constrained to the Quintet itself. Although we cannot exclude a contribution from the Milky Way, our analysis favours a magnetic structure within the SQ system. If indeed associated with the galaxy group in question, it would span a volume of at least $60\,\times\,40\,\times\,20\,\mathrm{kpc}$ and have a strength at least as high as that previously detected within large spiral galaxies. This field would then surpass the extent of any other known galactic, regular magnetic fields, have a considerable strength of a few microgauss, and would be the first known example of such a structure in a galaxy system other than a galaxy pair. Several other explanations are also presented and evaluated.

\end{abstract}

\section{Introduction}

	Magnetic fields are nearly omnipresent in the Universe: review studies (e.g. \citealt{beckrev}) clearly show that structures on all scales, from gas clouds in the interstellar medium (ISM) to galaxy clusters, are magnetised. Magnetic fields play an important role in physical processes that shape the evolution of galaxies and processes acting within, on a broad range of physical scales. Although the structure and importance of magnetic fields have been studied in a wide range of objects, galaxy groups have to date largely been ignored. However, as stated by \citet{hickson82}, the environment of a compact galaxy group is not calm, and this class of objects thus provides a unique view into numerous phenomena at the heart of galactic evolution, some of which are amplified by the mutual proximity of the individual member galaxies. It is therefore important to address our current lack of insight into the role and importance of magnetic fields structures on this scale.\\

    The most effective method to detect and study extragalactic magnetic fields is to observe the non-thermal radio continuum emission. Synchrotron radiation is produced when cosmic ray (CR) particles interact with magnetic fields. This emission is best observed in the radio regime. However, this technique also has some disadvantages when seeking to detect intergalactic magnetism. First and foremost, in the intergalactic medium (IGM) within and surrounding galaxy groups, the magnetic field is generally not directly illuminated due to a lack of massive stars that end their lives as supernovae, providing acceleration sites for fresh CRs. There are certain objects where vigorous intergalactic star formation does take place -- e.g. HCG\,7 and HCG\,22 (both studied by \citealt{torresflores09}), or HCG\,91 \citep{eigenthaler15} -- but these are compact star forming regions, not extended on the scale of haloes. The latter are usually devoid of any significant star formation. If there are no additional sources of electron re-acceleration, like strong shocks, the synchrotron spectrum steepens quickly (increasing values of $\alpha$ for a spectrum described by $S_\nu\propto\nu^{-\alpha}$, and deviations from the power-law curve), such that emission at higher frequencies fades out first. By itself this would not be much of a problem in massive objects such as galaxy clusters, where the total flux of a steep-spectrum source, like a relic or a halo, is high enough to allow detection of objects with spectral indices as steep as $\alpha\sim 4$  (see e.g. \citealt{slee01}). However, detection of steep-spectrum emission is more problematic for galaxy groups, where the total emissivity is much fainter. Observing at lower radio frequencies is also problematic, because the angular resolution decreases with increasing wavelength; for rather small objects -- as \citet{hickson92} lists, most of the compact groups he studied are not larger than a few arcminutes in diameter -- it is usually impossible to distinguish between individual galaxies, the IGM, and background sources. This situation is now changing with the advent of next-generation low-frequency radio telescopes such as LOFAR \citep{LOFAR}.\\
    
The magnetic field can be classified on the basis of its degree of ordering. Most of the detected fields are turbulent. There is no distinguishable structure on larger scales, as its direction changes rapidly from place to place. Such a field manifests as non-polarised -- i.e. total intensity signal only. If the field is at least weakly ordered, there is also a polarised component, which can usually be detected in Stokes Q and U (which together describe linear polarisation). The ordered fields have two subtypes; which particular one is found depends on the mechanism that caused the ordering. Those which are ordered, but not regular, are created e.g. when a turbulent field is compressed due to a passage of a shock front or stretched by a large-scale shearing flow. As a result, there is an apparent common orientation, but it is not kept over the larger scale, as the sign can change from point to point. The other subtype -- genuinely regular magnetic fields -- are unidirectional, and an effective regularising mechanism is required to produce a field that has consistent orientation and sign over a large scale. This can be usually achieved by the action of the magneto-hydrodynamic dynamo process (see \citealt{beckrev} and references therein). As these two subtypes indicate the presence of very different phenomena, it is crucial to effectively distinguish them. It is usually achieved by the analysis of Faraday Rotation associated with the observed sources. Passage of an electromagnetic wave through an ionised medium that hosts magnetic field introduces rotation of its plane of polarisation. This change in the angle of polarisation is proportional to the wavelength squared, the strength of the magnetic field inside the medium, and its sign. In the case of a regular magnetic field, the sign is constant among the line of sight (LOS), so the result is non-zero overall rotation. In the case of ordered (and turbulent) fields, the sign tends to change many times along the LOS -- and as a result, on scales larger than those typical for turbulence (an outer scale of a few kiloparsecs in case of galaxies, see eg. \citealt{mao15}), the overall Faraday rotation averages down to zero. The slope of the relation between the observed rotation angle and the wavelength squared is called the Rotation Measure (RM): it depends on the strength and directionality of the field, its extent among the LOS, and the thermal electron content.\\

The easiest way to estimate the RM for a particular, polarised radio source is to analyse data from multifrequency observations. Owing to the fact that the rotation angle is proportional to $\lambda^{2}$, it takes at least three independent frequency bands to begin to extract this information unambiguously. Analysis of the polarised angle distribution yields information about the total induced Faraday Rotation, which enables calculation of the RM (and, along with some extra assumptions, the regular magnetic field component that is parallel to the LOS). However, the only information gained that way is the total RM: if there is a number of media that each introduce different Faraday Rotation, it is not possible to disentangle their individual contributions. Thus, observations should be conducted at a broader range of frequencies. The possibility of mapping emission in Faraday space was first explored by \citet{burn66}, who introduced the concepts of Faraday Depth (FD), and Faraday Dispersion Function (FDF). The FD -- often confused with the RM -- is the 'intrinsic RM' of a particular volume that the radiation is passing through. As there may be many of them along the LOS, the observed RM is the sum of the contributions from all of them. The FDF is basically a functional relation that describes how the polarised surface brightness depends on FD. Burn's pioneer work has later been extended by presenting a method to recover the FDF even in cases of a (highly) incomplete sampling of the $\lambda^2$ space -- the so-called RM Synthesis technique \citep{brentjens05}. As the RM-Synthesis utilises an analogon to the Fourier Transform, reverting from the wavelength space to that of the Faraday Depth results in the emergence of sidelobes in the telescope beam analogon -- the Rotation Measure Spread Function (RMSF). And exactly as the \textsc{clean} algorithm \citep{clean} deals with deconvolution in case of the incomplete sampling of the $(u,v)$ plane, its 'Faraday version', the \textsc{rm-clean} \citep{sings2} allows to effectively minimise the unwanted implications of a limited sampling in the Faraday domain.\\

	One of the best studied galaxy groups is HCG\,92, otherwise known as the Stephan's Quintet. This system is the first described galaxy group, named after its discoverer, \'Eduard Jean-Marie Stephan \citep{stephan77}. It is believed to be a compact triplet that has encountered a close passage with two other galaxies. Interaction with the first of them is implicated in the formation of tidal tails -- both gaseous and stellar -- that are generally interpreted as clear signs of ongoing interactions \citep{moles98}. The system also hosts a large, intergalactic shock front that is believed to be a result of an ongoing collision of the group's IGM with NGC\,7318B, the current intruder \citep{osullivan09}. The shock front was first detected by \citet{allen72}, at 1.4\,GHz. The gas-deficient galaxy NGC\,7319 is located eastward of the shock. Its gaseous content was nearly entirely swept away during the previous interactions, forming the tidal tails \citep{moles98}. The other two members of the Quintet, the ellipticals NGC\,7317, and NGC\,7318A, are less pronounced. The picture is complemented by NGC\,7320, a dwarf interloper galaxy, that was long believed to be a member of the Quintet, the past intruder, NGC\,7320c, and as many as 20 tidal dwarf galaxy (TDG) candidates  \citep{hunsberger96}. This abundance of galaxies is not expected and creates quite a dense group environment -- the formation of TDGs is much more efficient in loose systems of field galaxies \citep{kaviraj12}.\\
    
    In the past years, the Quintet has been a popular target for radio continuum emission studies (e.g. \citealt{kaftankassim74, gillespie77, vonkapherr77}). The most recent work was done by \citet{xu03} and \citet{bnw13b}. Both of these papers used VLA interferometric data at 1.4 and 4.86\,GHz (the latter study also added single-dish data at 8.35\,GHz). The former aimed to provide high-resolution imaging to study the shock region, while the latter sought to reveal extended emission, possibly uncovering any signs of ordered magnetic fields (which are detectable via linearly polarised continuum emission). Taken together, these papers draw a picture with a plethora of radio-emitting structures: not only is the shock region visible, but the whole group is immersed in a large, magnetised structure, that spans an area of more than 100$\times$100\,kpc, and subsumes all member galaxies apart from NGC\,7317 (NGC\,7320c is also not included, but it has most probably left the vicinity of the Quintet, \citealt{moles98}). Inside the newer tidal tail, a local maximum of radio flux density marks the position of the TDG candidate SQ--B  \citep{xu03,bnw13b}. Its non-thermal spectral index confirms that it is magnetised; along with the intergalactic starburst region SQ--A in the same system, they comprise two out of three examples of TDG with detectable magnetic fields known so far. Most interesting in the context of the present paper is the large size of the area where polarised emission is detected: most of the group, including the shock region and certain areas between the member galaxies is visible. Moreover, there are hints that this field, at least in certain areas, is genuinely regular \citep{bnw13b}. 
Regular fields are sometimes found to have been decoupled from their original host galaxies, and typically follow tidal structures such as in the case of the Antennae \citep{chyzy04}. This situation has never been found in galaxy groups, where collisions between multiple members of a system could quickly disrupt the fine structure of such a field. Stephan's Quintet is the ideal case to search for regular fields in the environment of a galaxy group -- by the means of analysing the RM.\\
    
    In this paper we present the results of recent radio observations of Stephan's Quintet, aimed at revealing the possible large-scale regular magnetic fields lurking inside the IGM of this group. The paper is organised as follows: basic properties of the instrument, data, and the methods used to build the Faraday Spectra are described in Section~\ref{sec_data}. The Total Power (TP) map, and Polarised Intensity (PI) distributions at various Faraday Depths (FD), are presented in Section~\ref{sec_res}. Section~\ref{sec_disc} encapsulates the discussion of the results and their possible interpretations; there we present a confirmation of the existence of an intergalactic magnetic field associated with the Quintet. In Section~\ref{sec_conc}, we summarise our findings and conclude. The main dataset presented in this paper was acquired using the Westerbork Synthesis Radio Telescope -- the very same instrument which was used for the first radio study of the Quintet, nearly 50 years ago \citep{allen72}.
    
\begin{figure*} 
%\vspace{3.5cm} 
\resizebox{\hsize}{!}{\includegraphics{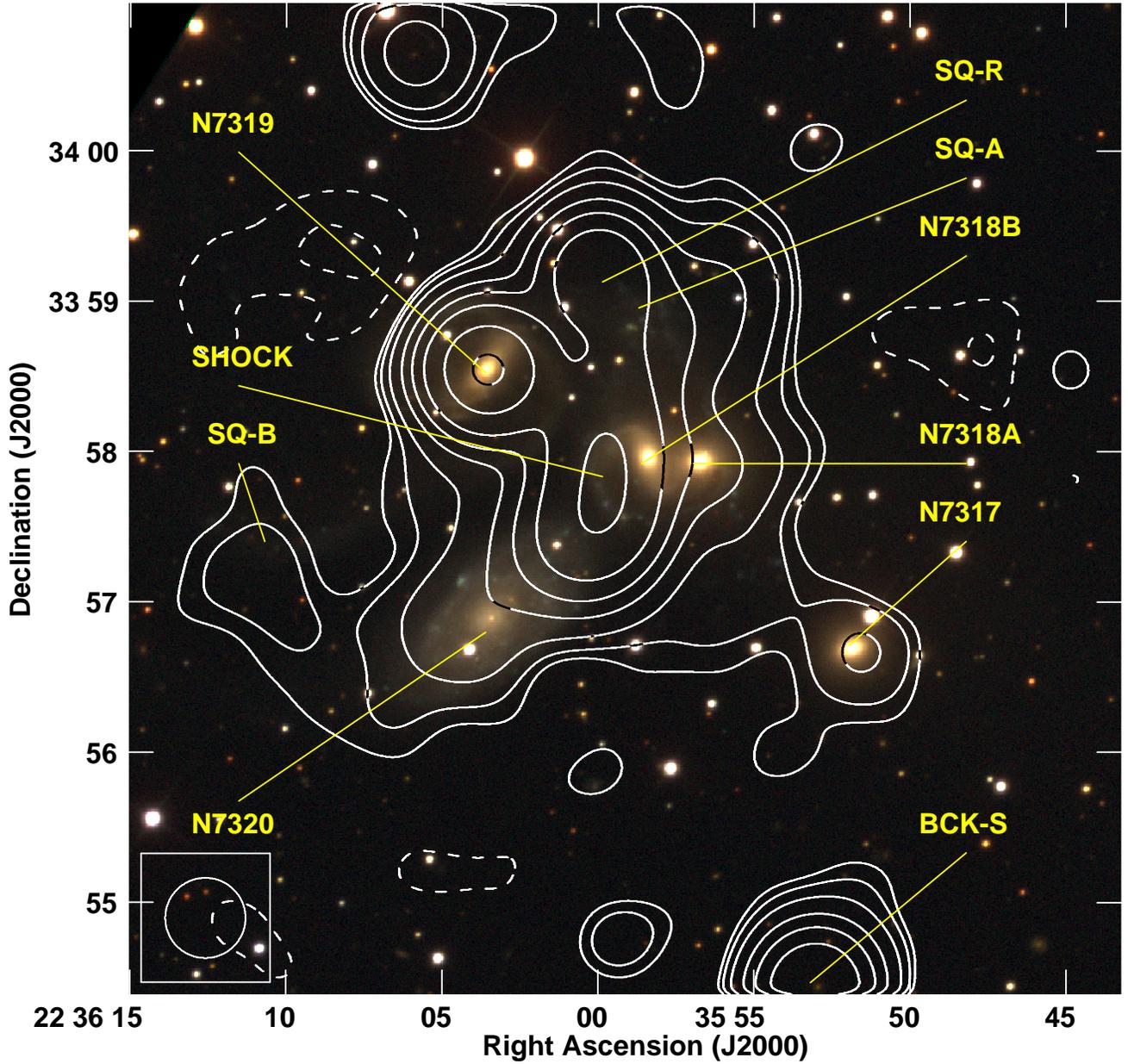}} 
\caption{\label{tp_full} Map of the Total Power emission of the Quintet at 1381 MHz, overlaid on a pseudocolour RGB composite of the SDSS irg maps. The angular resolution is 31 arcseconds and is illustrated by an ellipse in the bottom-left corner. The contour levels are --5, --3, 3, 5, 10, 25, 50, 100, 250, 500 $\times$ 0.05\,mJy (r.m.s. noise level) Designations of the sources referenced in the text are also included.} 
\end{figure*}

\begin{table*}
\caption{\label{tab:zdata} Right ascension, declination (both J2000), and redshifts for the sources analysed in this study}
\begin{center}
\begin{tabular}{lccccc}
\hline
Source			& R.A. (J2000)	&	DEC (J.2000)    &	    z	           &   Remark                 & Reference for z	\\
\hline
NGC 7317  		& $\mathrm{22^h35^m52^s}$       & $+33^{\circ}56'42"$& 0.02201 $\pm$ 0.00009&  SQ                   & \citet{hickson92} 	\\
NGC 7318 A		& $\mathrm{22^h35^m57^s}$	    & $+33^{\circ}57'56$"& 0.02211 $\pm$ 0.00008&  SQ                   & \citet{hickson92} 	\\
NGC 7318 B		& $\mathrm{22^h35^m59^s}$	    & $+33^{\circ}57'57$"& 0.01926 $\pm$ 0.00008&  New intruder         & \citet{hickson92}	\\
NGC 7319		& $\mathrm{22^h36^m04^s}$	    & $+33^{\circ}58'33$"& 0.02251 $\pm$ 0.00001&  SQ                   & \citet{nishiura00} \\
NGC 7320		& $\mathrm{22^h36^m04^s}$	    & $+33^{\circ}56'53$"& 0.00226 $\pm$ 0.00007&  Foreground interloper& \citet{hickson92}	\\
NGC 7331		& $\mathrm{22^h37^m04^s}$	    & $+34^{\circ}24'56"$& 0.00272*             &  Nearby galaxy        & \citet{sulentic01}	\\
Shock ridge		& $\mathrm{22^h36^m00^s}$       & $+33^{\circ}57'30$"& 0.01869 -- 0.02262** &  SQ, IGM shock front  & \citet{sulentic01}	\\
SQ--A   		& $\mathrm{22^h35^m56^s}$	    & $+33^{\circ}59'20"$& 0.02232***           &  SQ, TDG candidate    & \citet{sulentic01}	\\
SQ--B   		& $\mathrm{22^h36^m10^s}$	    & $+33^{\circ}57'22"$& 0.02207***           &  SQ, TDG candidate    & \citet{sulentic01}	\\
SQ--R   		& $\mathrm{22^h36^m00^s}$	    & $+33^{\circ}59'12$"& No optical id. known &  Background source    & 	\\
BCK--S   		& $\mathrm{22^h35^m53^s}$	    & $+33^{\circ}54'53"$& No optical id. known &  Background source    & 	\\
N1  		    & $\mathrm{22^h36^m06^s}$	    & $+34^{\circ}10'49"$& No optical id. known &  Background source    &	\\
N2  		    & $\mathrm{22^h35^m55^s}$	    & $+34^{\circ}14'16"$& No optical id. known &  Background source    &	\\
\hline
\end{tabular}
\end{center}
*) Uncertainty not provided;\\
***) \citet{sulentic01} provides a range of velocities at which IGM in the vicinity of the shock region was observed;\\
****) Calculated from velocities given by \citet{sulentic01}.
\end{table*}

\section{Observations and data reduction}
\label{sec_data}

The WSRT data for the Stephan's Quintet were acquired under the project code R13B011 in September 2013. A total of 24 hours of observing time were allocated; two 12h-long runs were employed, one centred at 1366\,MHz, and the other at 1697\,MHz. Each of these spectral windows had a total bandwidth of 160 MHz, divided into eight subbands of 64 channels each. To provide the highest possible sensitivity to the weak, extended structures, the instrument was arranged in its \textit{Maxishort} configuration. The radio galaxy 3C286 was chosen as a calibrator for amplitude, bandpass, and phase. \\

The pre-processed datasets have been imported into the Astronomical Image Processing System (\textsc{aips}) following a standard procedure outlined in the WSRT users manual. Erroneous data points -- e.g. RFI contamination -- have been removed following manual inspection and flagging. Polarisation information has also been corrected (especially the conventions for defining the directions on the Poincar\'e sphere). The $(u,v)$-data were imaged and deconvolved using the \textsc{imagr} task to produce the final total intensity map. They were $(u,v)$-tapered to obtain a circular beam of 31\arcsec. The data were self-calibrated in a number of loops, starting from the phase-only scheme, and ending with phase and amplitude corrections simultaneously derived. The solution interval was chosen to gradually decline from 10\,minutes down to 1 with the increasing quality of the subsequent steps. Corrections from the last loop were then applied to the $(u,v)$-data and the final set of maps was created: an averaged Stokes I map, and a number of channel maps for the Stokes Q and U parameters -- all with a correction for the primary beam shape applied. This was done using a fifth order polynomial, with the coefficients provided by K. Kreckel of the MPIA\footnote{http://www.mpia.de/homes/kreckel/wsrtpbcor.html}. 

Rotation Measure (RM) Synthesis was then conducted on the linearly polarised data. To suppress the sidelobes resulting from incomplete $\lambda^2$ sampling, we used the \textsc{rm-clean} \citep{sings2} algorithm to deconvolve the RMSF from the reconstructed Faraday Depth spectra.  These steps were first carried out using our own software, and the results were subsequently cross-checked with the \textsc{pyrmsynth}\footnote{https://github.com/henrikju/pyrmsynth} code as described in Section \ref{sss:ambiguity}. In case of both approaches, several Faraday cubes with different ranges and stepsizes in Faraday Depth were generated to ensure that the analysis is robust to numerical issues (e.g. avoiding issues from an improper choice of the parameters related to RM-deconvolution). The final image sets used in this study were made with a range of $-$1500 to +1500 \radmm\ and a RM step of 10 \radmm. The resolution of the final data cubes (for Q, U, polarised intensity, and polarisation angle) in the Faraday Depth space is 150\,\radmm.

Throughout the paper we adopt the nomenclature that all names applying to Faraday space, as adaptations of common astronomical vocabulary (e.g. sidelobes of the FWHM of the main lobe of the RMSF) will be preceded by a prefix ``RM-'', e.g. RM-cleaning, RM-sidelobe etc. to minimise confusion.

\subsection{Definition of the Stokes QUV parameters}
\label{altdefs}

The final definition of the Stokes parameters (convention) was chosen by the IAU \citet{iau74}, after the WSRT was built; the reference system of the WSRT is different from the currently used one\footnote{https://www.astron.nl/radio-observatory/astronomers/analysis-wsrt-mffe-data/analysis-wsrt-mffe-data} and usage of the software that was made for a different telescope -- like \textsc{aips}, made for the VLA -- could result in misidentified Stokes parameters, and/or an error in their signs. In addition, \textsc{aips} defines them using the RR and LL polarisations, while WSRT -- using XX and YY. This also needs to be accounted for. To test whether the adopted conversion relations are correct, 3C286 -- which was used as the dataset's primary calibrator -- was imaged in all four Stokes channels and the results were compared with the literature data\footnote{http://www.astron.nl/radio-observatory/astronomers/analysis-wsrt-data/analysis-wsrt-dzb-data-classic-aips/analysis-wsrt-d}. It turns out that, with the exception of the (true) V signal, for which the sign can not be evaluated (due to 3C286 having a near-zero circular polarisation, resulting in a nearly null value of this parameter), the derived values are consistent with expectations from the literature, up to 1\% in case of Stokes I, and up to 10\% in case of Stokes QU (See Table \ref{tab:caldata}). 

\begin{table}
\caption{\label{tab:caldata} Values derived from imaging the calibrator data compared to the WSRT manual information}
\begin{center}
\begin{tabular}{lcccc}
\hline
Source			&	I	&	Q	&	U	&	V	\\
\hline
Uncorrected data		& 14.50	& -1.36	& 0.02	& -0.55	\\
WSRT manual	    & 14.65	&  0.56	& 1.26	&  0.00	\\
Converted data  & 14.50 &  0.55 & 1.36  & $\pm$0.02 \\
Conversion* 	&	I	&	-V	&	-Q	& +/-U	\\	
\hline
\end{tabular}
\end{center}
*) These are the channels and factors used to obtain the final channel QU data.
\end{table}

\section{Results}
\label{sec_res}

\subsection{Total Power}

The map of the Total Power (TP) emission from the Quintet, presented in Fig.~\ref{tp_full}, reveals an extended radio envelope covering all the original group members, as well as the interloper galaxy NGC\,7320, the current intruder NGC\,7318B, the TDG candidates SQ--A and SQ--B, and large areas of optical void at the south-eastern and north-western edges of the radio structure. Some basic information on the sources referenced in the text is provided in Table~\ref{tab:zdata}. NGC\,7317 is also visible, with a narrow (about one beam in width), bridge-like connection to the main structure. The total flux of the envelope (with NGC\,7317) is $91 \pm 5$\,mJy. The highest peak flux density is attributed to the core of NGC\,7319, and is equal to $27 \pm 2$\,mJy/beam. South from the group, several background sources were detected; they are, however, much weaker than the Quintet, with the most luminous one among them having a peak flux of $11 \pm 1$\,mJy/beam. The angular resolution does not allow us to separate the contribution from SQ--A from the surrounding radio emission.

\subsection{Polarised Intensity}

%=============================
\begin{figure}
    \includegraphics[width=0.45\textwidth]{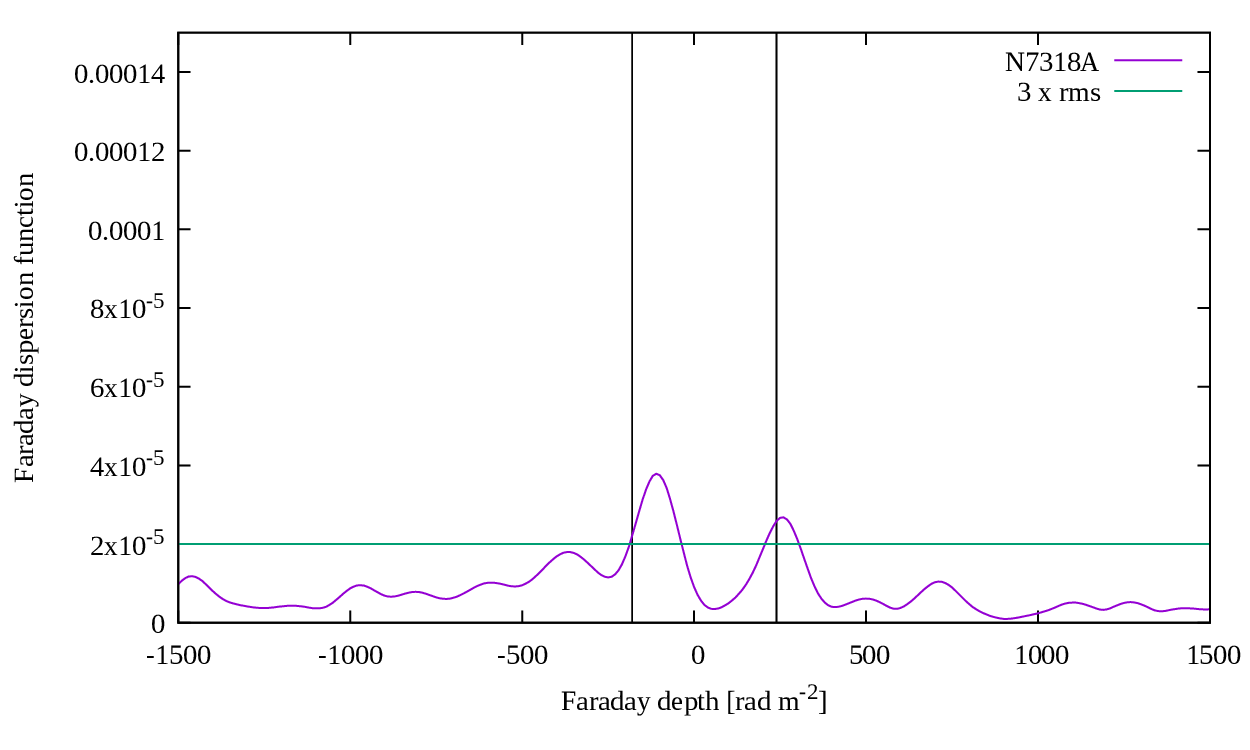}
    \includegraphics[width=0.45\textwidth]{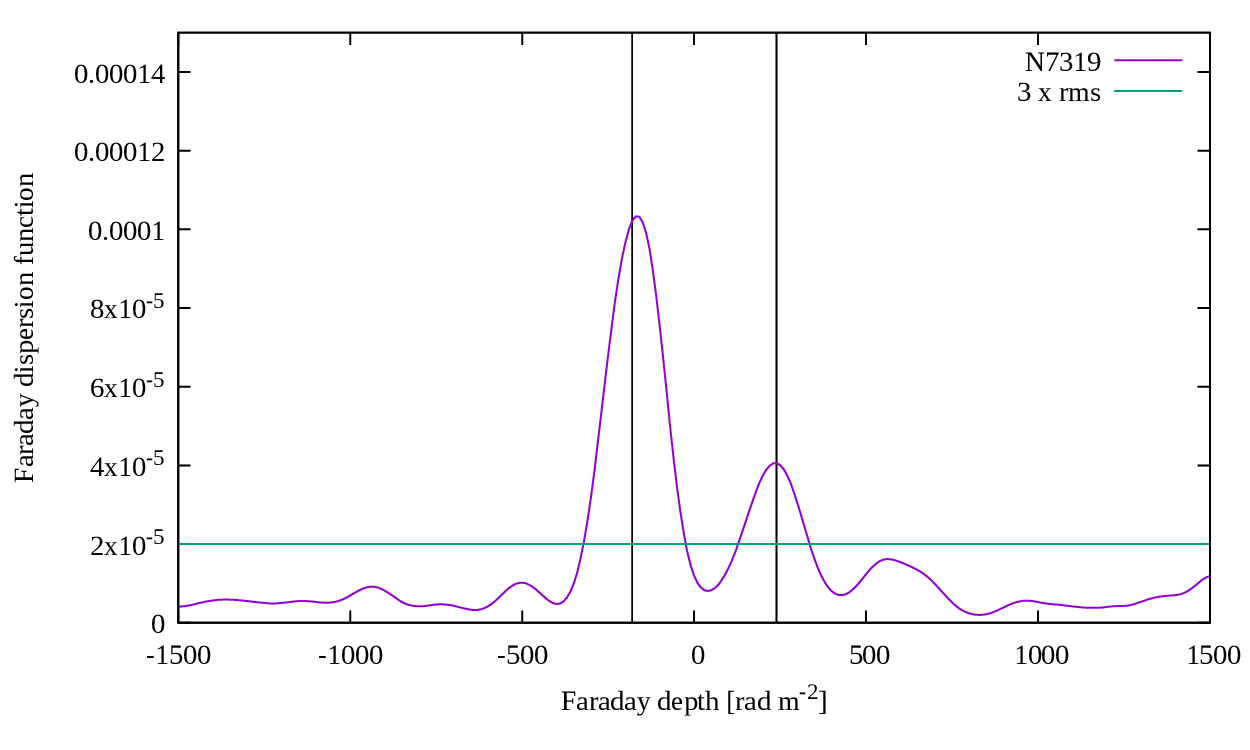}
    \includegraphics[width=0.45\textwidth]{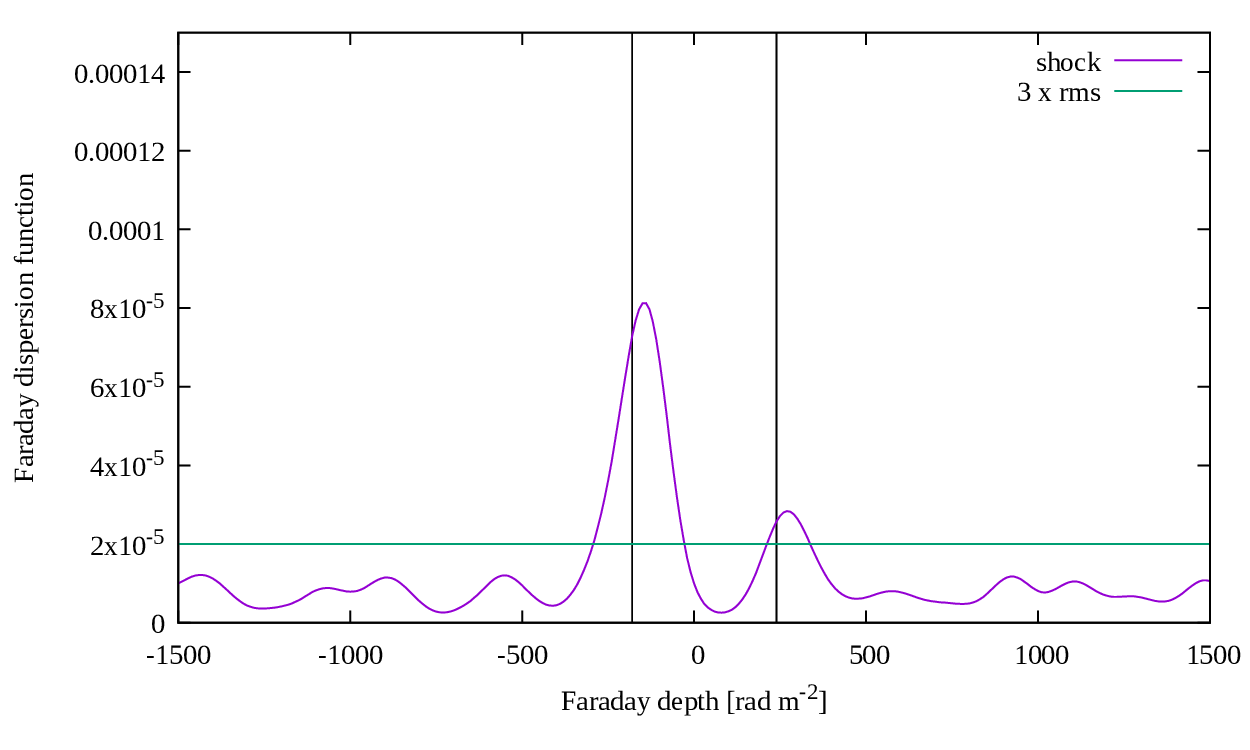}
    \includegraphics[width=0.45\textwidth]{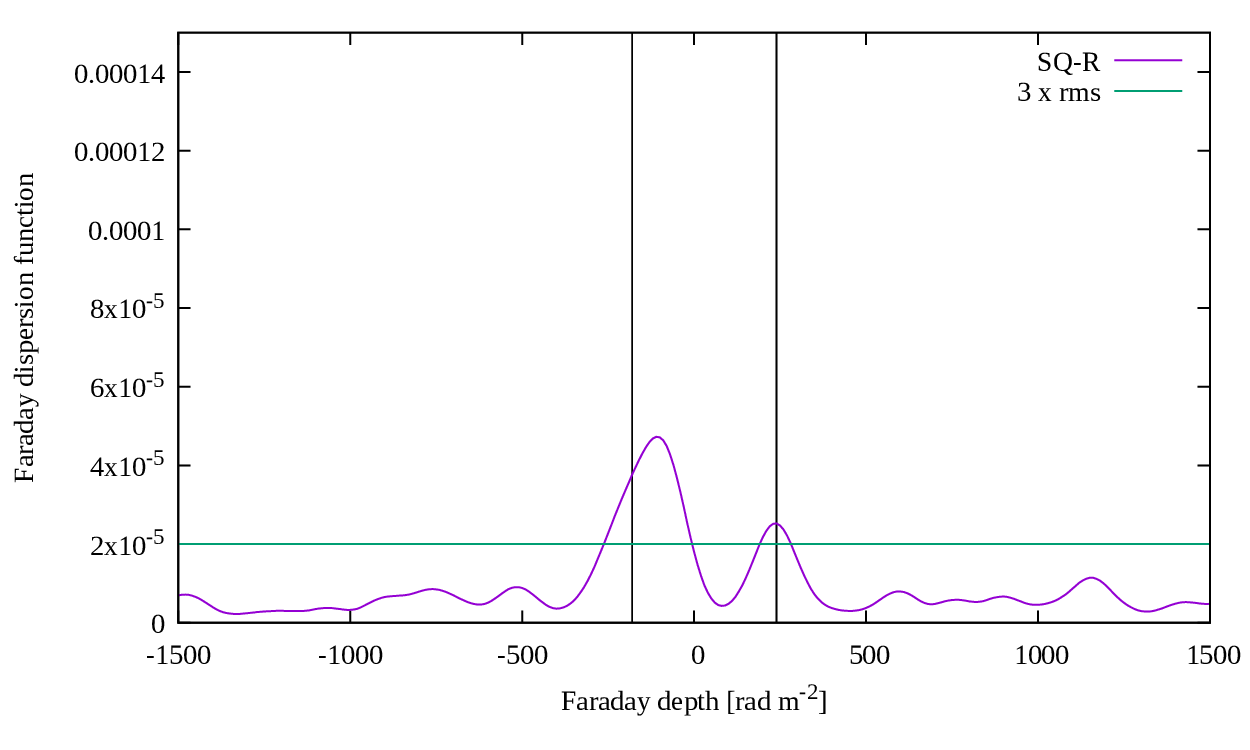}
    \caption{Faraday Dispersion Function (FDF) plotted for the sources inside the area of the Stephan's Quintet. The amplitude axis is scaled in Jy/beam/RMSF. From the top to the bottom: NGC\,7318A, the core of NGC\,7319, IGM shock region, and background source SQ--R. The green line signifies the 3$\times$\,r.m.s. level at the center of the map of 20$\mu$Jy/beam, while the vertical black ones are used to point out the --180 and +240 \radmm{} depths.}
     \label{pi_prof_sq}
\end{figure}
%=============================

%=============================
\begin{figure}
	\begin{center}
    \includegraphics[width=0.45\textwidth]{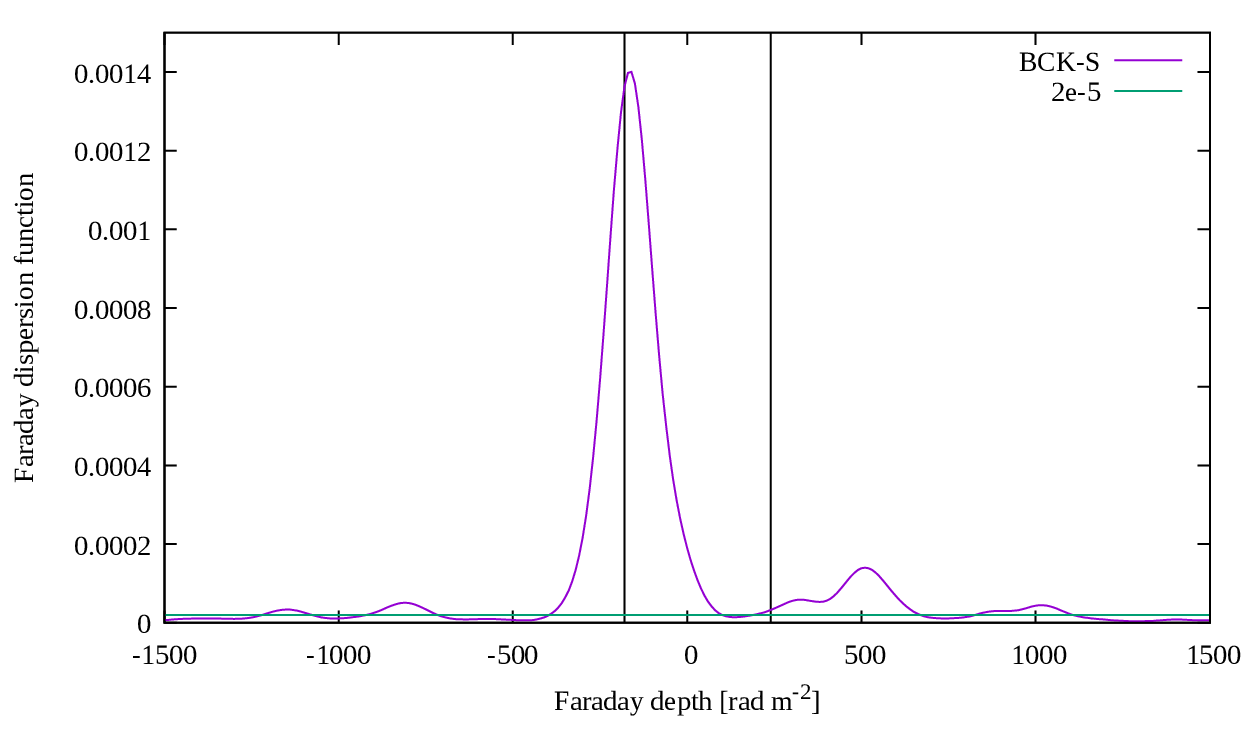}
    \includegraphics[width=0.45\textwidth]{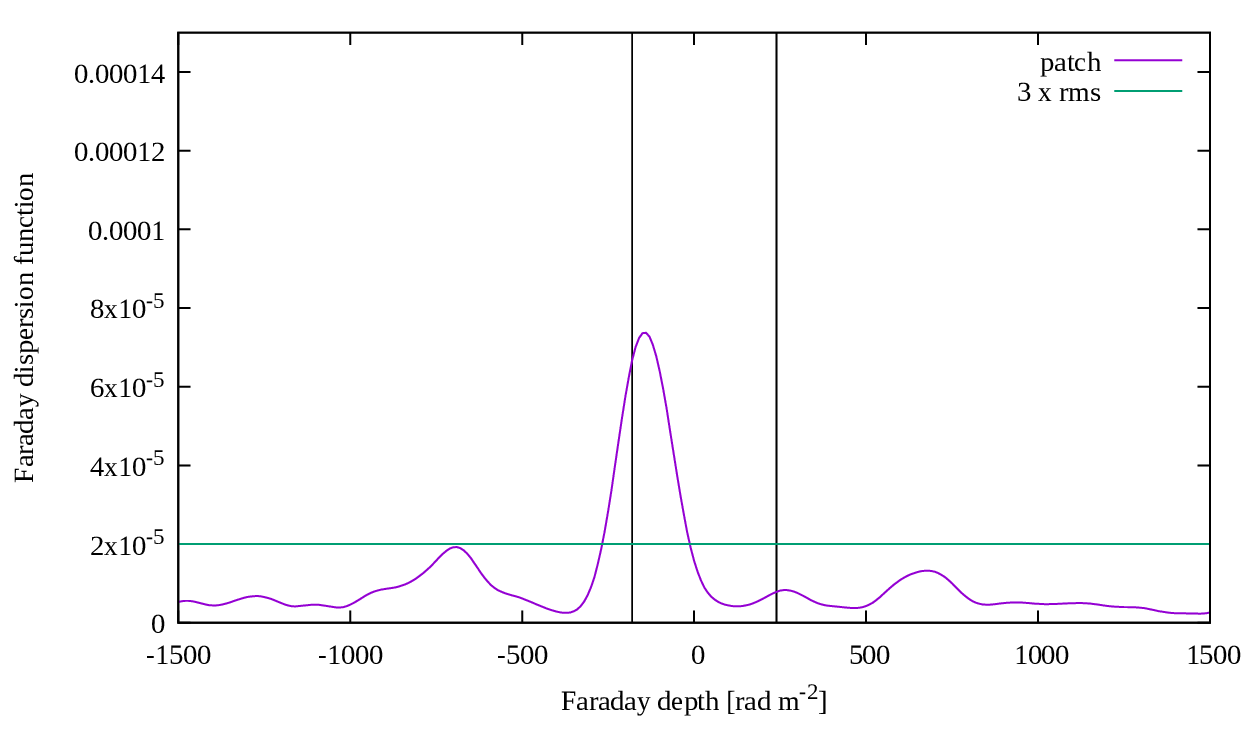}
    \includegraphics[width=0.45\textwidth]{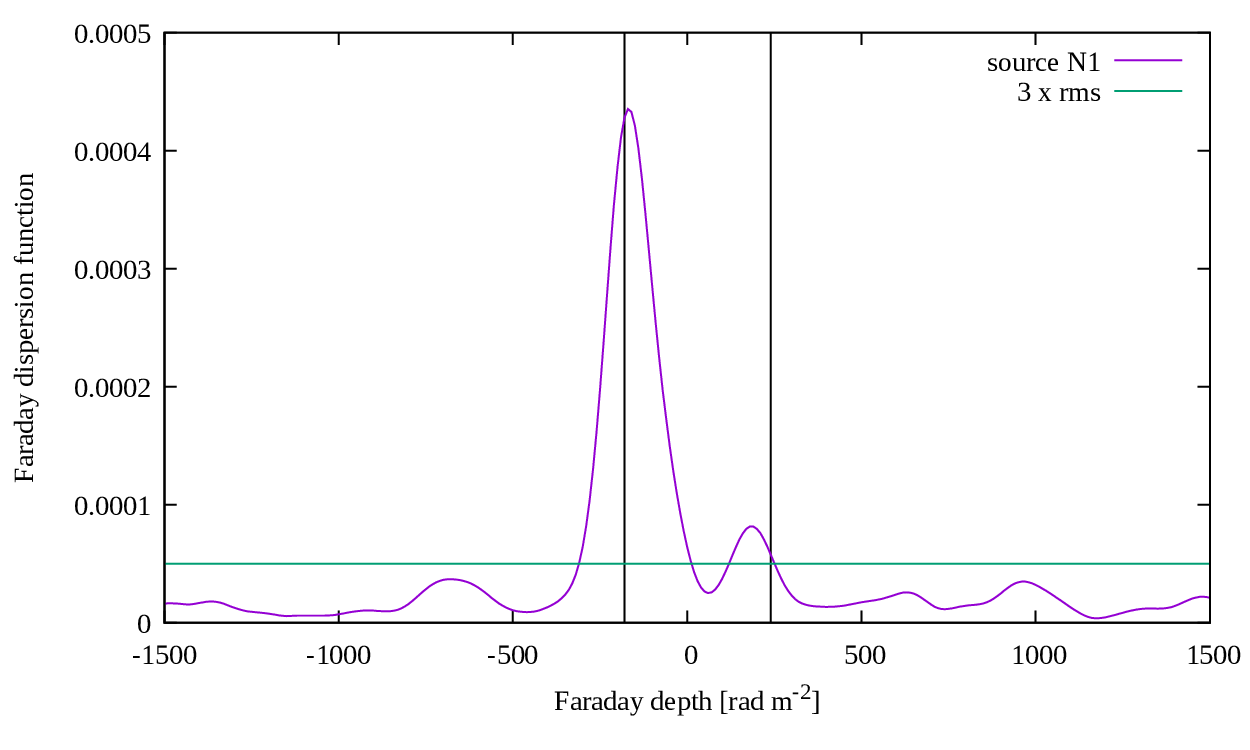}
    \includegraphics[width=0.45\textwidth]{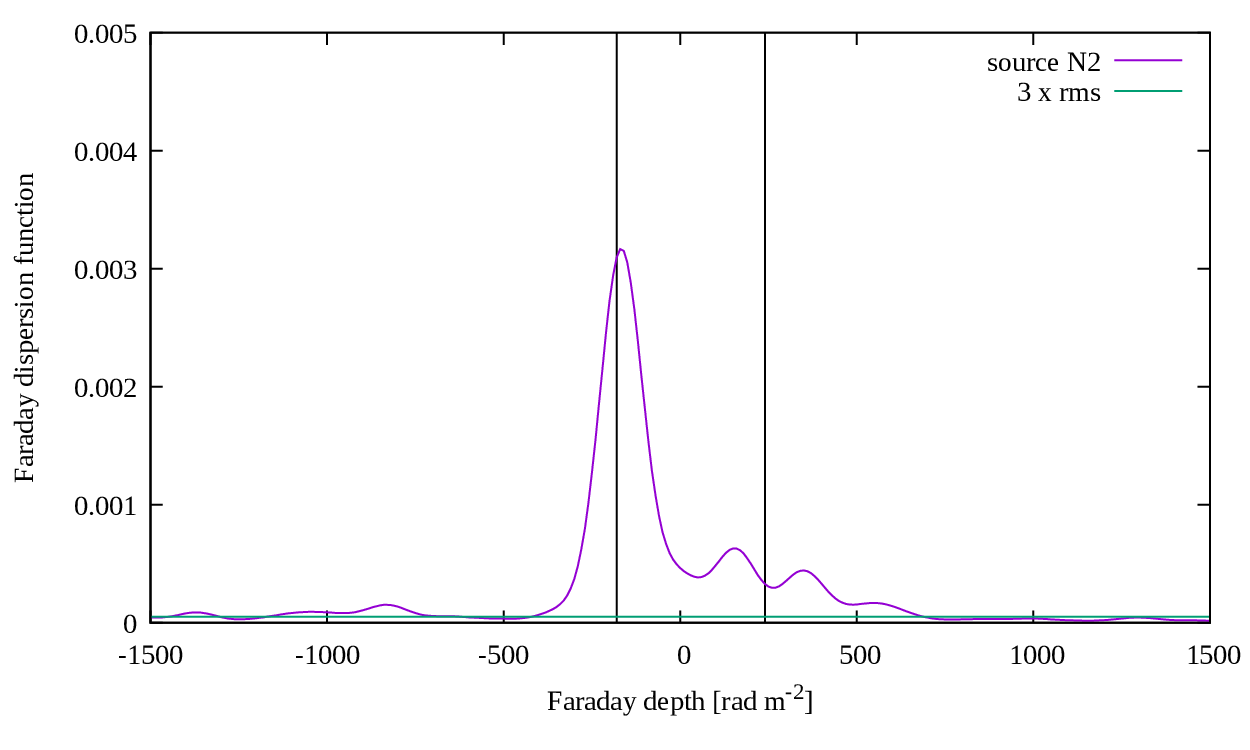}
    \caption{Faraday Dispersion Function (FDF) plotted for the comparison sources. The amplitude axis is scaled in Jy/beam/RMSF.  From the top to the bottom: background source BCK-S, a sample from the area covered by a large patch of Galactic emission, background source N1, background source N2. The green line signifies the local 3$\times$\,r.m.s. level (upper panels -- 20$\mu$Jy/beam, lower panels -- 50$\mu$Jy/beam), while the vertical black ones are used to point out the --180 and +240 \radmm{} depths. Note the lack of the +240\,\radmm{} peaks in these profiles, in contrast to those shown in Figure~\ref{pi_prof_sq}.}
     \label{pi_prof_off_sq}
	\end{center}
\end{figure}
%=============================

The polarised data product is a cube with as many as 300 individual planes, each representing a different Faraday Depth. The full, imaged field of view is approximately 1 degree in diameter, well beyond the half-power point of the primary beam and including regions where strong contamination from instrumental polarization is expected. The most convenient way to present the results is to show a set of maps at different depths (and positions) where significant emission was detected as well as profiles presenting the Faraday Dispersion Function (polarised intensity as a function of Faraday Depth), extracted from particular areas on the map. The thin, green, horizontal line visible in the profiles presented in Figures~\ref{pi_prof_sq} and~\ref{pi_prof_off_sq} signifies the assumed 3\,r.m.s. level: 20$\mu$Jy/beam for the sources in the central part of the map, and around 50$\mu$Jy/beam for those close to its boundaries. The r.m.s. values were estimated as an average of the r.m.s. of the individual planes after RM-Synthesis (before primary beam correction). The solid, vertical black lines are used to illustrate the Faraday Depths of specific interest: $-180$ and +240\,\radmm.

\begin{figure*}
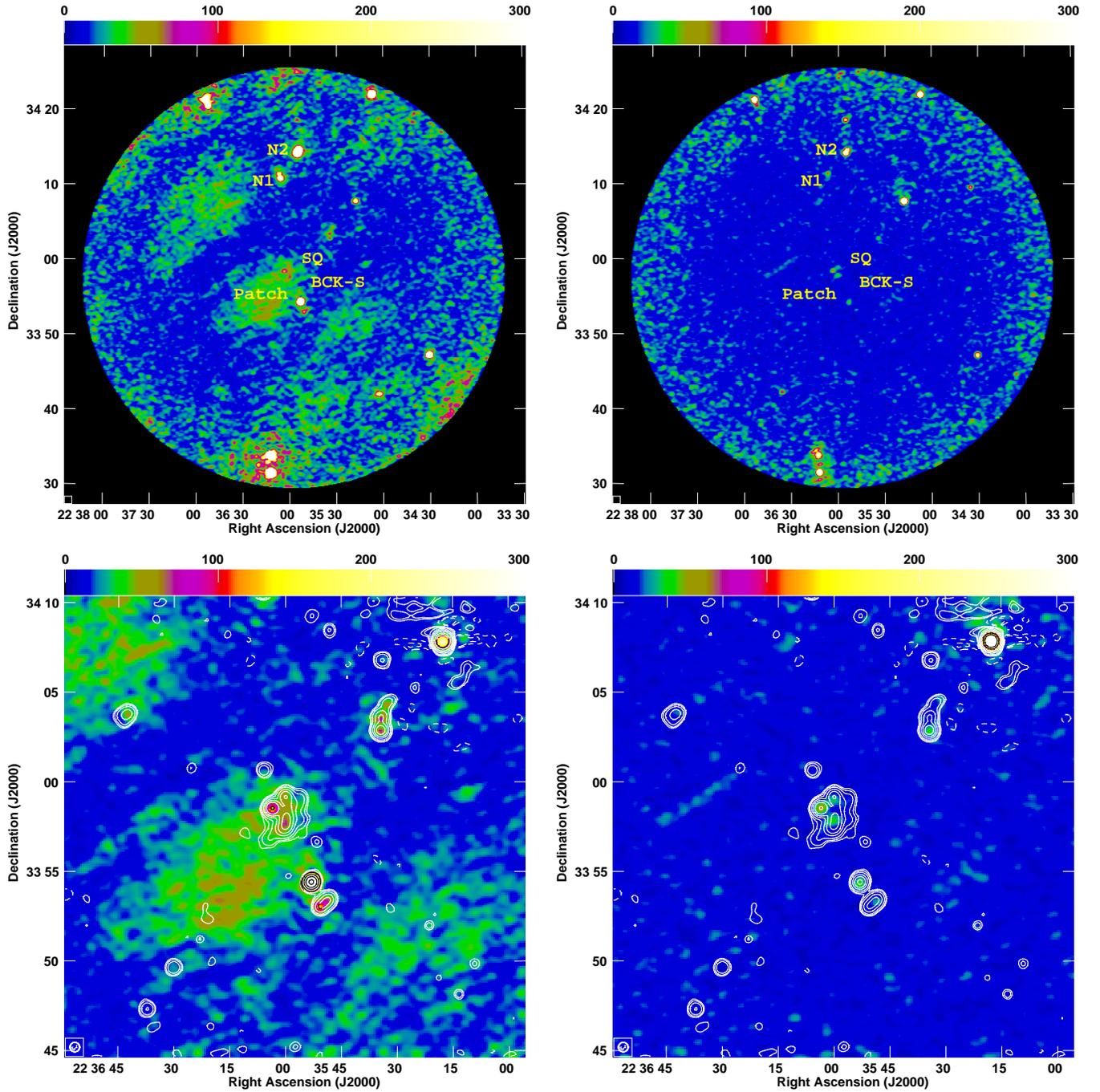

%\vspace{3.5cm} 
\includegraphics[width=0.50\textwidth]{PI-180_nocont.PS} 
\includegraphics[width=0.50\textwidth]{PI+240_nocont.PS}
\includegraphics[width=0.50\textwidth]{SQ-180_CUTOUT.PS} 
\includegraphics[width=0.50\textwidth]{SQ+240_CUTOUT.PS}
\caption{Emission at $-$180 (left panels) and +240 \radmm (right panels) in pseudocolours. The upper panels show the whole FOV of the WSRT, while the lower ones are limited to the area where instrumental effects in the Faraday space are subdominant, and are overlaid with contours of the Total Power distribution. The angular resolution is 31 arcseconds. The large, unlabelled structures visible in the field at $-$180\,\radmm\ are interpreted as diffuse Milky Way emission. The colour wedge is scaled in units of $\mu$Jy/beam/RMSF. The contour levels are --5, 3, 5, 20, 100, 200, 500, 1000, 2000, 5000 $\times$ 0.1\,mJy/beam. Notice that these levels are different from those in Fig.~\ref{tp_full} so that noise signal enhanced by the primary beam correction far from the centre of the map is suppressed.  The apparent peaks at the locations of sources N1 and N2 are actually the result of an elevated baseline level in the Faraday spectra there; see Fig.~\ref{pi_prof_off_sq}.} 
\label{field_pi}
\end{figure*}

Visual inspection of different planes along the Faraday Depth axis reveals layers where large (> 2') linearly polarised structures can be easily identified. The bulk of these emitting regions is located near FD = $-$180\,\radmm\ (Fig.~\ref{field_pi}, left panels and can be naturally associated with the foreground (Milky Way) magnetic field; such an inference is drawn on the basis of the Galactic Rotation Measure studies from \citet{taylor09}, \citet{oppermann12}, \citet{oppermann15}, and \citet{hutschenreuter20}. All these works suggest that the median value in the vicinity of the Quintet is equal to approximately $\rm -160$ to $\rm -180$\,\radmm -- fairly consistent with the values derived from the RM Synthesis. Areas contributing to the Galactic emission can be seen mostly northwards from the phase centre, overlapping the Quintet. The Faraday Depth associated with the position of the peak flux density tends to vary slightly when measured in different positions; the range of the variability is $-$130 to $-$190\,\radmm. The typical uncertainty associated with peak Faraday depths measured from RM cubes is $\sigma_\mathrm{FD}=\mathrm{FWHM}/(2\cdot \mathrm{S/N})$ where FWHM is the width of the RMSF (150\,\radmm\ as indicated above), and S/N is the signal-to-noise of the associated peak \citep{brentjens05}. For many lines of sight in our cube, the main peak has a S/N around $5-10$, so we expect typical $\sigma_\mathrm{FD}\approx8-15$\,\radmm. Thus a substantial contribution to the observed spread in Faraday Depth values is likely due to measurement uncertainties.\\

\begin{figure}
%\vspace{3.5cm} 
\resizebox{\hsize}{!}{\includegraphics{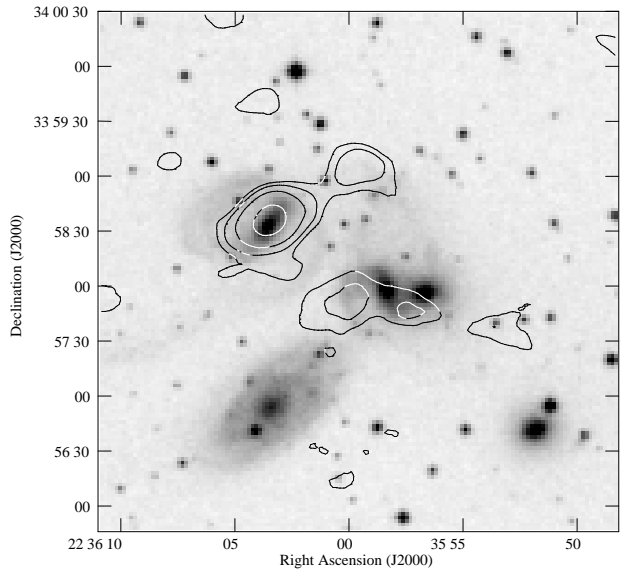}} 
\caption{Contours of the emission at +240 \radmm overlaid on a POSS-II-F (red) map. The angular resolution is 31 arcseconds. The contour levels are 3, 4, 6, 10 $\times$ 6.3\,$\mu$Jy/beam/RMSF (r.m.s. noise level).} 
\label{sq_pi}
\end{figure}

Fig.~\ref{pi_prof_sq} shows the Faraday Dispersion Function (FDF) for the objects that belong to the Stephan's Quintet: NGC\,7318A (uppermost), NGC\,7319 (upper-middle) and the IGM shock region (lower-middle). The lowermost panel presents the profile extracted for the background radio galaxy SQ--R \citep{xu03} These four profiles show a double-peaked distribution. One of the peaks, around $-$180\,\radmm\, was already associated with the Galactic contribution. The second one, at around +240 \radmm\, was not attributed to any previously mentioned structure, suggesting that it might be intrinsic for the Quintet. Other sources in the field seem not to exhibit a similar spectral configuration (see Fig.~\ref{pi_prof_off_sq}). This is also confirmed by looking at the map of polarised emission at $+$240 \radmm, where the Quintet is the only bright object near the centre of the map (Figs.~\ref{field_pi}, right panels, and Fig.~\ref{sq_pi}). Faraday Depth spectra plotted over the lines of sight in the angular vicinity of the Quintet (so still covered by the large structure at $-$180\,\radmm), but remote enough so no emission associated with the group can be present there (Fig.~\ref{pi_prof_off_sq}, 2nd profile) also show no emission at that depth, indicating that it is not associated to the already detected, Galactic structure.

Throughout the Quintet area, the position of the second peak remains quite stationary: it reaches +230\,\radmm\ at the position of NGC\,7319, +250\,\radmm\ at the position of SQ-R, +260\,\radmm\ at the position of NGC\,7318A and the highest value of +275\,\radmm\ is found in the shock area. In each of these cases, the peak intensity of both peaks exceeds the 3\,$\times$\,r.m.s.\, level (after primary beam correction). For S/N=3, one would expect a nominal RM uncertainty of $\approx20$\,\radmm. Thus the location of the second peak is remarkably stable, and differences appear to be consistent with expected measurement errors.

In the case of the off-SQ profiles, no matter the distance from the Quintet, and the direction, there is no emission peak at that depth. While the origin of the emission visible in the spectrum of BCK--S is unknown, the raised signal level at around +240\,\radmm\ seen in its profile is only a ``tail'' of a peak centred at around +320\,\radmm. The +500\,\radmm\ peak visible over the 3$\times$\,r.m.s. line is likely an effect of standing-wave interference (see Sec.~\ref{wave_int} for the discussion of this effect), prominent for this source in particular because it is considerably brighter in Stokes $I$ than those in Fig.~\ref{pi_prof_sq}.
Secondary peaks in sources N1 and N2 are detected with low confidence due to the large angular distance of both sources from the pointing center (both are beyond the half-power point of the primary beam for the higher of the two frequency bands, and in the case of N2, for both frequency bands). At these field offsets, polarization fractions $\lesssim\,1\%$ have a high probability of being instrumental in origin \citep[e.g.,][]{dbb05}, but due to strong frequency dependence and rotational asymmetries \citep[see][]{popping_braun_2008} it is difficult to quantify this more precisely, particularly given the broad frequency coverage of our observations.
All together, we conclude that the MW peak near $-180$\,\radmm\ is reliable in all sources illustrated in Figs.~\ref{pi_prof_sq} and \ref{pi_prof_off_sq}; that the peak at 240\,\radmm\ is reliable only for sources in the SQ region; and that the low significance and field locations of the sources exhibiting additional peaks indicate that they are subject to instrumental artefacts.

Among the additional Faraday Depth cubes that we produced, one was especially made to search any possible emission at high values of FD. This cube had range of $\pm$20000\,\radmm\ and consisted of 1000 planes separated by 20\,\radmm. No clear signal at extreme Faraday depths was detected.

\section{Discussion}
\label{sec_disc}

\subsection{What structure does the regular magnetic field belong to?}

	The method of revealing a regular magnetic field by reconstructing the Faraday Dispersion Function relies on the fact that a genuinely regular magnetic field is characterised by a non-zero RM value; a large-scale, coherent magnetic field along the line of sight (LOS) induces a cumulative sense of Faraday rotation, whereas in the cases of random or ordered fields, the incoherent orientation of the field averages down to RM=0 on scales relevant for this study, and no overall rotation is observed. Given the large separation in Faraday Depth space between the detected polarisation components in the lines of sight to the members of Stephan's Quintet, we can identify the field (or fields) detected along the LOS to the Quintet as being regular. The key issue when discussing them is whether they can be associated with the group, or are in the foreground. This can be attempted through the analysis of the Faraday spectra of the sample sources. All objects that have been detected and identified as parts of the Quintet show a double-peaked distribution with a foreground peak at $-$130 to $-$180\,\radmm, and another peak at +230 to +275\,\radmm. The fact that such peaks appear at slightly different depths -- and that a shift of the position of the first peak is not correlated to the the shift of the second, and vice versa -- is another argument in favour of the real character of the detected structures.\\

    As the emission at around +240\,\radmm\ seems to be morphologically associated to the area of the SQ -- contrary to the much more spatially extended emission at the characteristic Faraday Depth that is associated with the Galactic foreground -- this indicates that the observed emission originates within or around this galaxy group. As each of the detected sources that belong to the Quintet reveals the presence of this feature, the rotation must happen somewhere within or in the front of the group. There are no traces of a large-scale structure throughout the map at the depth of +240\,\radmm; it is even visible -- in the spectrum of NGC\,7318A -- that while the amplitude of the signal at $-$180\,\radmm\ drops down near the edge of the large patch, this does not happen in case of the $+$240\,\radmm\ peak, that has a constant strength measured for all the sources in the area of the Quintet. All other sources which appear in the map at the FD of +240\,\radmm\ have maxima at different depths, are affected by instrumental leakage at levels consistent with their secondary Faraday peaks, and thus are not connected to the Faraday structure that we associate with the Quintet. What is important in case of the radio galaxy SQ-R is that the background emission travelling through the area of the Quintet is not altered by any other medium -- at least not significantly.  Apart from the two main peaks, no other significant emission has been detected at higher positive and negative depths for the sources within the SQ.
    
    If the detected magnetic field is indeed associated with the IGM of the Stephan's Quintet, this means that the Faraday rotation induced by this medium is equal to the difference between its internal FD and the FD of the next source in the line of sight. Therefore, the RM induced by the Faraday screen is then of approximately 350--500\,\radmm.

\subsection{Alternative hypotheses}

	The claim of the existence of regular magnetic field associated with the Quintet (and derivation of its properties) is based on the relative and absolute position of the two peaks that were detected in the Faraday space. To further test the conclusion presented above, we performed various tests, and analysed several different hypotheses: they include both technical, and physical considerations, and they are summarised in subsections listed below.
      
\subsubsection{Different scale, or offset in the Faraday Depth}

One of the most obvious alternative interpretations is that there is an artificial offset/scaling factor in the Faraday Depth of our data. However, relevant large-scale studies -- Faraday sky maps by \citet{taylor09}, \citet{oppermann12}, \citet{oppermann15}, or \citet{hutschenreuter20} as well as the study of the nearby galaxy NGC\,7331 done by \citet{sings2} -- consistently locate the Galactic foreground in this region at a depth of about $-160$ to $-$180\,\radmm\, concordant with our results. Thus, we further disfavour any scenario in which the detected Faraday Depth signals are artificially shifted or scaled by some amount, on the basis that such an effect would also cause the Galactic peak to be shifted away from the local value identified by other researchers (which is visibly not the case).

\subsubsection{RM-sidelobe amplification of noise}

\begin{figure}[H] 
%\vspace{3.5cm} 
\resizebox{\hsize}{!}{\includegraphics{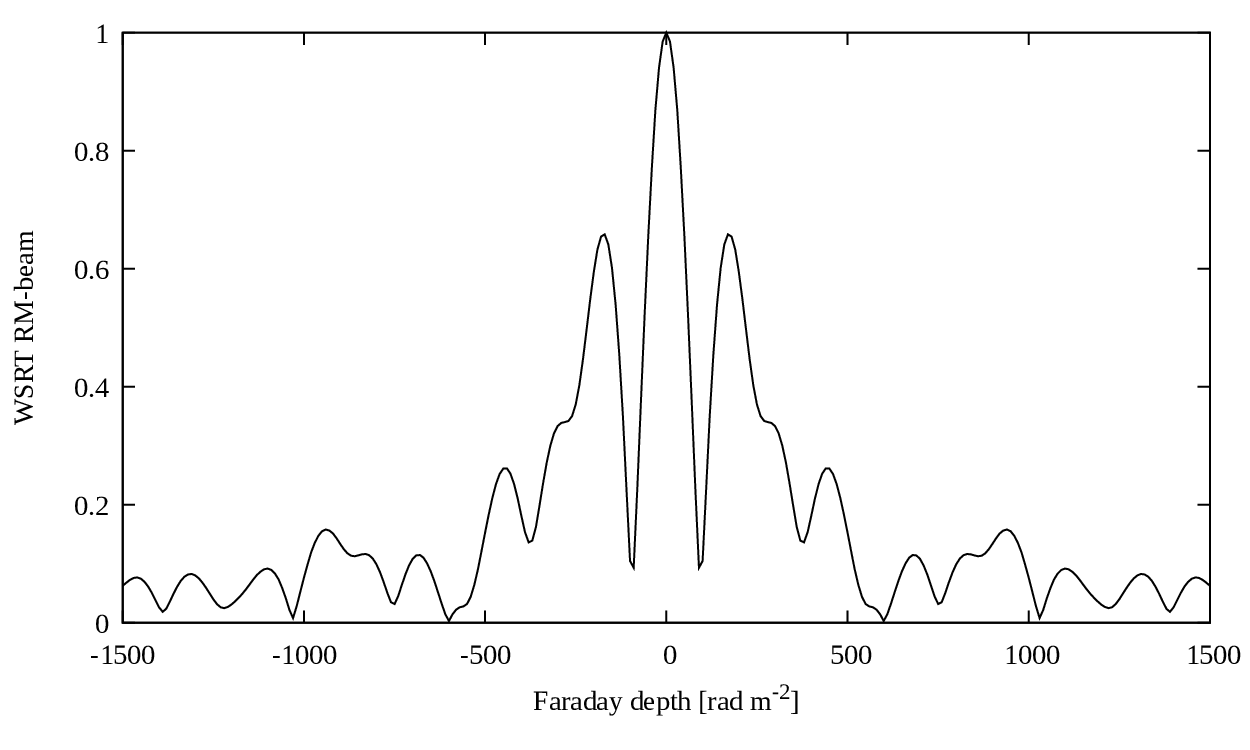}} 
\caption{RMSF derived from the WSRT data} 
\label{rmtf}
\end{figure}

Another natural explanation of the detected second peak is that it is just an RM-sidelobe corresponding to the more luminous, Galactic peak; all sources in the Quintet have lower intensity, and are located considerably far from the ``main'' one. The RMSF profile can be seen in Fig.~\ref{rmtf}, and the positions and signal values for the first four sidelobes (relative to the main peak) are listed in Table~\ref{tab:rmtf}. First of all, should the +240\,\radmm\ peak be a RM-sidelobe of the $-$180\,\radmm\ one, there should also be a mirrored one at about $-$600\,\radmm; this does not happen. There is also no evidence of either the first, or the second sidelobe of the main peak. Also, the relative distance between the two peaks varies, and the range of variations (350--420\,\radmm) falls between the sidelobes -- close to the local null of the RMSF. Furthermore, the ratio of the peaks is variable (between about 1.5 to 3.0) and inconsistent with the levels of the sidelobes of the RMSF. Last but not least, the RM-dirty map was also investigated, showing the presence of RM-sidelobes of the Galactic signal in positions consistent with the RMSF shape, but not in the location occupied by the +240\,\radmm\ peak. Instead, the +240\,\radmm\ peak shows its own RM-sidelobes, present at the positions consistent with the RMSF shape. If this signal was an RM-sidelobe of the Galactic peak, it would not have any RM-sidelobes of its own.

\begin{table}[H]
\caption{\label{tab:rmtf} Positions of the sidelobes of the RMSF, and their intensities (relative to the main peak)}
\begin{tabular}{p{23mm}ccccc}
\hline
Sidelobe					&	Main	&	1	&	2*	&	3	&	4	\\
\hline
Absolute depth (\radmm)	&	0		& 170	& 290	& 440	& 680	\\
Relative signal				&	1.00	& 0.66	& 0.34	& 0.26	& 0.12	\\
\hline
\end{tabular}\\
$*$) The position and peak signal of the 2$\rm^{nd}$ sidelobe are ambiguous due to its immersion in the first one. 
\end{table}

\subsubsection{Ambiguity of the RM-Synthesis}
\label{sss:ambiguity}

Studies done by \citet{sun15} suggest that different implementations of the RM-Synthesis technique yield slightly different results, and the output of different algorithms may significantly vary in case of the presence of multi-component, or Faraday-thick slabs. As already mentioned in Section~\ref{sec_data}, to ensure that what we see is a fair representation of the actual structure in Faraday Depth space, calculations were repeated using the \textsc{pyrmsynth} package (which handles the data in a fundamentally different way, e.g. through the use of a Fast Fourier Transform instead of a Discrete Fourier Transform). The results differed at negligible levels with respect to the spectrum shape, peak position, amplitude of the signal, and RA/Dec position. Additionally, we performed RM-Synthesis for NGC\,7331, and compared the results with the archival ones from WSRT-SINGS \citep{sings2}, which was calculated using the code by \citet{brentjens05}. Owing to the fact that this galaxy lies sufficiently far from the pointing centre of our WSRT dataset, only about 2\,\% of its original flux is recovered due to the primary beam response. Hence, the absolute intensities of the peaks that represent Milky Way and NGC\,7331 are lower in our data; however, the relative intensities, as well as obtained Faraday Depth are consistent with our results. 
We could not detect the second peaks in the spectra of sources N1 and N2. Either they are real, but the sources themselves are too far from the pointing centre of that observation, or they are indeed instrumental in character.

\subsubsection{Standing wave interference}
\label{wave_int}
\begin{figure}[H] 
%\vspace{3.5cm} 
\resizebox{\hsize}{!}{\includegraphics{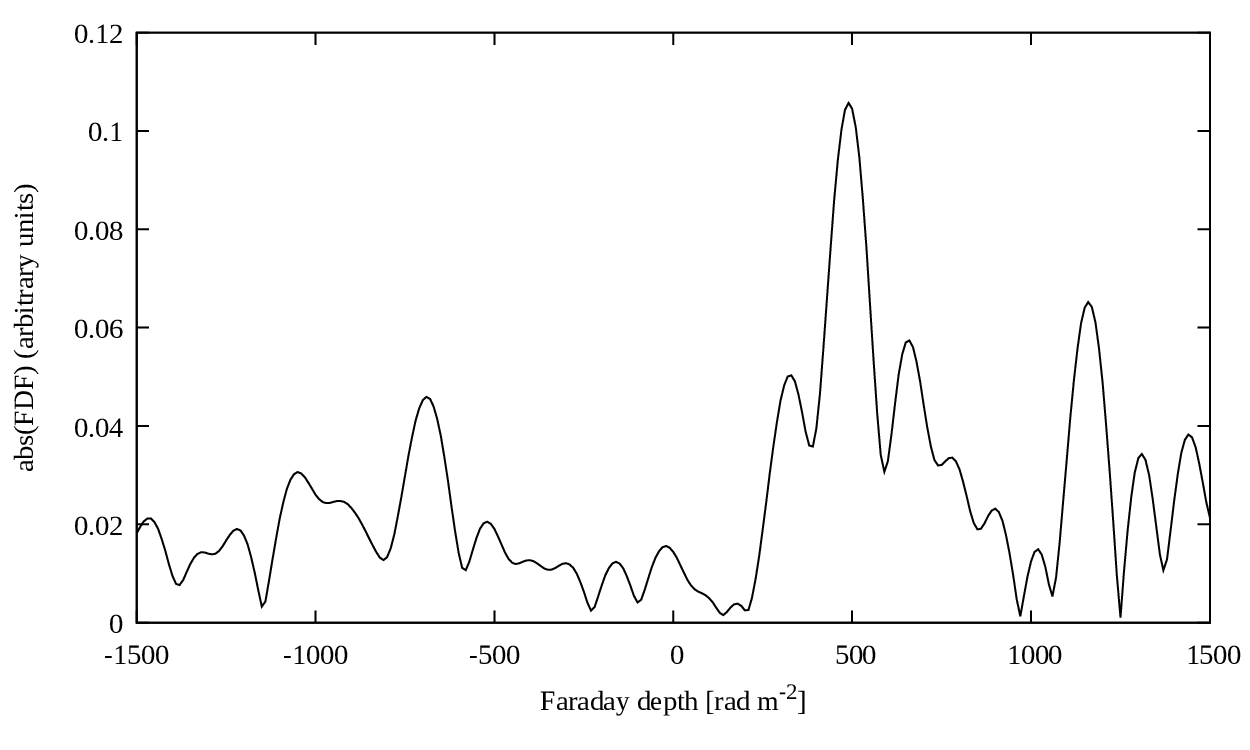}}
\caption{Illustration of the effect of a standing wave interference in the data. If present, the 17\,MHz modulation would create an artificial peak at the FD of $\approx$+500\,\radmm\ .} 
\label{fig:17mhz}
\end{figure}

Observations done with the WSRT can be subject to an unwanted contamination resulting from the generation of standing waves between the dish, and the receiver. This effect has a frequency of 17\,MHz, and can be partially mitigated using special calibration techniques (which are not implemented by us for this work). There is a possibility that it would be represented in the Faraday spectrum; however, calculation of the expected modulation of the polarised intensity profile due to the presence of standing waves  (Fig.~\ref{fig:17mhz}) suggests that the response caused by this effect does not correlate with the ones that were detected for the sources inside the area of the SQ. There is a possibility that the spectrum of BCK--S (Fig.~\ref{pi_prof_off_sq}, uppermost panel) shows such a peak; however, this is the brightest object in our PI maps (more than an order of magnitude stronger than any of the sources inside the Quintet) and is certainly not associated with the SQ. Halving, or doubling its frequency in search for harmonics produces spectra that are inconsistent with the ones obtained. We conclude that standing-wave effects possibly induce the secondary peak in the spectrum of BCK--S, but have no other substantial impact on our data.\\

\subsubsection{Faraday thick structure confined to the Milky Way, or two Galactic structures}

The maximum observable scale in Faraday Depth space (before substantial depolarisation occurs) depends on the highest frequencies included in the data \citep{brentjens05}. In case of the SQ dataset, this characteristic value is approximately 110\,\radmm, so any extended Faraday structure would be discretely sampled as a series of peaks (or just the two boundary ones). Therefore, what has been assumed to be a peak related to the medium inside the Quintet could be just a boundary of a Faraday-complex structure that spans from app. $-$200\,\radmm\ to +250\,\radmm, and located inside our Galaxy. Another possibility could be that the two peaks do not belong to one structure, but are two distinct regions both within the Milky Way.
However, we consider both of these scenarios to be unlikely. First of all, those sources and structures that do not overlap the Quintet  show a single reliable, narrow peak at $-$130 to $-$190\,\radmm. Association of the second peak exhibited by sources in the SQ region with structure in the MW would imply a sudden transition from a Faraday thin medium to a Faraday thick medium, exactly at the location of the SQ. No other sources in the field show two peaks with the same characteristic RM difference.

Compounding the issues described above, another problem arises if one tries to reconcile the data solely with sensible values for the Galactic regular magnetic field. The literature suggests that the latter is typically around 1.5\,$\mu$G (\citealt{beckrev}, estimated from pulsar RMs); the electron density inside our Galaxy can be assumed to be equal to around 0.1\,$\mathrm{cm}^{-3}$, which is the average value for the warm, ionised medium inside the ISM \citep{ferriere01} -- we are not aware of any data in the literature that allows us to constrain this value.
These numbers lead to the conclusion that the minimal pathlength to keep the resulting $\mathrm{B_{REG}}$ at a level similar to that estimated by \citet{beckrev} would be around 2--3\,kpc. To further investigate the reliability of this scenario, we have searched the literature for models of the Galactic magnetic field in this region, and compared our results with those of \citet{vaneck10} and \citet{han18}. These models consistently suggest that in the direction of the Quintet (l=93$^{\circ}$), the regular magnetic field is relatively weak and does not show any sign reversals. Also, while the LOS is pointed across the Perseus and Outer arms, only one magnetised structure seems to be detected. Therefore, we cannot attribute each of the detected peaks to separate spiral arms -- they would have to originate in a single structure. However, there are no signs of a sign reversal on the way -- yet such an feature would be required to explain why both peaks have been detected with different signs, suggesting opposite directions of the magnetic field. While the strength of the field may be locally enhanced, it would be difficult to reconcile existing models of the structure and directionality of the Milky Way's magnetic field with our observed peaks at negative and positive Faraday depth. Last but not least, the expected ``thickness'' of a spiral arm is around 800\,parsecs from one boundary dust lane to another \citep{vallee14}. This leads to a contradiction with the minimal expected pathlength, twice or even thrice.
While we cannot rule out either variation of the scenario where the Milky Way is responsible for all of the polarized emission in the field, we consider it to have a lower likelihood than the picture wherein the emission at $+$240\,\radmm\ originates from the SQ. Polarimetric observations focused on achieving a narrower RMSF, and with improved sensitivity and angular resolution, would be needed to definitively rule out this scenario.

\subsubsection{A foreground, but extragalactic structure}

The last possibility considered here is that the +240\,\radmm\ peak, although real and extragalactic, originates in a structure that is not connected to the Quintet: it could be located in the foreground of the group, such that the sources inside the Quintet shine through. Based on the known geometry of the group, the natural candidate for a host of such a structure is NGC\,7320. However, its optical boundaries lie much further to the South than any of the objects that show the extra peak. Also, the neutral gas distribution \citep{williams02} does not extend that far, and the galaxy's emission itself is unpolarised. This altogether suggests, that NGC\,7320 is not the host of the regular field. In this case the only possibility that is left is that there is a detached magnetised cloud, somewhere in the void space between the Milky Way, and the Quintet, and similar to the latter's angular size. Such a scenario would be highly fine-tuned.

\subsection{Properties of the magnetic field detected in the IGM of the Stephan's Quintet}
\label{sq_MF}

We conclude that the most likely scenario is that the detected regular field is associated with the Quintet itself. The next questions are: how strong is the regular field, and what is its extent? The textbook formula that relates these two quantities with the rotation measure, as given by \citet{brentjens05} is presented below:
\begin{equation}
\phi\,[\mathrm{rad/m^2}] = 0.81 \int B_{||}\,[\mu\mathrm{G}]\times n_{e}\,[\mathrm{cm^{-3}}]\times dl\,[\mathrm{pc}]
\end{equation}
i.e. the Faraday Depth ($\phi$) is the thermal electron density ($n_e$) weighted integral of magnetic field component ($B_{||}$) along the line of sight ($l$). The $n_{e}$ parameter can be estimated from e.g. X-Ray emission. As an upper limit to its value, we use the mean electron density in the shock region taken from \citet{osullivan09} -- 1.167$\times 10^{-2}$\,$\mathrm{cm^{-3}}$. However, we expect the Faraday rotating medium to be placed at a considerable distance from the shock itself, where $n_e$ should be lower. In order to establish a lower limit for the electron density, we use Figure 6. from \citet{osullivan09}, describing the dependence between X-Ray surface brightness (which is proportional to the electron density squared), and the distance from the centre of the group. We estimate the ``background'' electron density -- at the distance of more than 40\,kpc from the centre -- as approximately 0.5$\times 10^{-2}$\,$\mathrm{cm^{-3}}$. Therefore, the product of regular magnetic field strength (in $\mu$G) and pathlength (in kpc) should take a value of 50 -- 100 [$\mu$G$\times$kpc].

We can estimate the maximum strength of the regular component of the magnetic field based on the results of \citet{bnw13b}, who estimate the average total strength of the magnetic field in the Quintet as approximately 6\,$\mu$G, and up to 11\,$\mu$G in the shock area.
Therefore, we can constrain the total magnetic field strength to at most a few $\mu$G, and thus the pathlength distance to 10--40\,kpc, depending on the electron density.

It is important to recognise that the bulk of the regular magnetic field cannot be constrained to be located within the shock region: in this picture, one would likely detect each of the individual objects (shock, NGC\,7318A, NGC\,7319, SQ-R) at different Faraday Depths. It is then apparent that what is revealed by the Faraday data cube is a magnetic screen, through which all polarised components of the group are observed; a magnetised envelope that covers at least the northern half of this system. Assuming the linear scale of 0.442\,kpc/arcsec (based on \citealt{hickson92} and assuming H$_{0}$=73 km/s/Mpc), one can estimate the transverse extent of the magnetised structure to be at least 60$\times$40\,kpc in size. As there are no obvious reasons for the medium to abruptly become de-magnetised just beyond the boundaries outlined by these objects, the exact extent of the Faraday screen is likely somewhat larger. Nevertheless, even if considering only the quadrilateral area denoted by the four objects detected in the PI distribution, it is, by far, \emph{the largest magnetised structure found in a low mass galaxy system.} Comparing this structure to any other similar one is not yet possible, as regular, intergalactic magnetic fields have not been detected in any other galaxy group (the study by \citealt{farnes14} revealed that their target groups were unpolarised at 612\,MHz). The most appropriate structures to compare to would be then the tidal tail of the Antennae, where the regular magnetic field spans over 20\,kpc \citep{basu17}, and the magnetised bridge of the Magellanic Clouds \citep{kaczmarek17}. The strength of the regular magnetic field inside the Quintet seems to fall between that of those two structures (around 8\,$\mu$G for the former, and 0.3$\mu$G for the latter). The regularity level of the magnetic field inside the tail of the Antennae is close to the theoretical maximum (more than 60\% -- \citealt{basu17}), and the regularity of the Quintet field can be quite similar. Another example of a similar structure would be the ``Taffy bridges'' \citep{condon93, condon02B}. While observational evidence cannot distinguish whether their magnetic fields are regular or ordered, simulations carried out by \citet{vollmer12} for the first of these systems suggest that these intergalactic structures might indeed be populated by a genuinely regular magnetic field.
%We emphasise that the calculations assume that the magnetic field has a ``depth'' comparable to that of the shock region.
We consider our adopted values to provide a minimum estimate for its strength and energy density. The magnetic field under consideration might be much more compact along the line of sight. In particular, it is possible that such a field is even immersed in a larger, turbulent structure. Should this be the case, however, the net value of 5\,$\mu$G over approximately 10\,kpc implies that the ``core'' field responsible for the change in the RM values has a much larger (local) strength.\\
The ``screen'' model is also an approximation, chosen as the simplest way to (roughly) determine parameters of the magnetic field that we interpret as being connected to the Stephan's Quintet. 
The Quintet has a complicated history of interactions, and there are different scenarios describing how it was formed. Previous interactions are known to have formed at least two tidal tails, and stripped most of the gas from NGC\,7319 \citep{moles98}. Several low surface brightness features exist in the vicinity of NGC\,7319, in particular the diffuse, reddish halo detected by \citet{duc18}. This suggests that baryonic counterparts to the supposed magnetic structure may be present in the SQ system. Magnetohydrodynamical simulations done by \citet{geng12} suggest that none of the models and scenarios for the formation and evolution of the Quintet can be dismissed on the basis of the existing data and results of the simulations -- only indications of preference of one model over another can be discussed. In particular, as neither the existence of the extended radio halo, nor the polarised emission from the Quintet was known prior to the study carried out by \citet{geng12}, this information could not have been implemented into their models. Nevertheless, even without the observational evidence, existence of magnetised structures in the Quintet's area was deemed a likely occurrence. Should one of such structures exist, or one of the LSB features detected by \citet{duc18} be regularly magnetised, it could match the ``screen'' perfectly. In addition, more sensitive observations with better angular resolution would be needed to further discuss the configuration of the magnetic field inside the group in question. Detailed, numerical modelling -- taking into the account results of our current study -- should also be applied. In particular, multi-band, interferometric observations, spanning L- through X-band, performed with the VLA, would be extremely desirable. This is because better surface brightness sensitivity is needed to map the diffuse emission; higher sensitivity would allow detection of a larger number of extragalactic background polarized AGN and thus obtain a more reliable sense for the consistency of the Faraday peaks across this patch of sky (and better constrain the extent of the patch); and better Faraday resolution will enable separation of those peaks in the Faraday spectra that are now smeared together.

\subsection{Other sources in the field}
   
\subsubsection{SQ--B}

One of the most interesting results from radio studies of the Quintet is the presence of a magnetised TDG candidate, referred to as SQ-B. It was first mentioned by \citet{xu03}, who connected two regions of vigorous star formation to isolated patches of enhanced radio emission. \citet{bnw13b} argued that analysis of the then available data favoured its identification as a TDG, not just a detached region of star formation. Moreover, the 4.86\,GHz radiopolarimetric data suggested a high polarisation fraction ($\approx$ 30\%), giving birth to the idea that inside this object, the magnetic field is amplified through e.g. dynamo mechanism. This claim was also supported by a steep radio spectrum, typical for non-thermal emission.\\

The WSRT data generally confirm most of these findings, additionally revealing a bridge between the dwarf and the whole SQ envelope. Interestingly, it does not follow the tidal tail of NGC\,7319 -- within which the dwarf is believed to have formed -- but is oriented towards south/southwest, where the interloper lies. It follows almost exactly the neutral hydrogen tail of the Quintet \citep{williams02}. Moreover,
the maxima of the continuum and line emission are consistent. This is all suggestive that both the magnetic field and neutral gas are being stripped out of the system through the same tidal structure, and the TDG candidate might not be associated with the optical tidal tail, but rather with the gaseous one. It is indisputable that the emission from SQ--B is predominantly non-thermal; the spectral index calculated using the WSRT and the older VLA data is equal to $0.7 \pm 0.1$ -- meaning a spectrum far too steep to be thermal-dominated. It is somewhat flatter than previously estimated, but still comparable to the ones provided by both \citet{xu03} and \citet{bnw13b}. The reason for this discrepancy lies most probably in a significantly higher sensitivity of the new WSRT data. The ``young'' spectrum of SQ-B is not unusual, for it is a vigorously star-forming object, one of the most efficient regions of star formation in the whole system \citep{xu05}. The magnetic field strength, recalculated using the same values as previously by \citet{bnw13b}, is $4.5 \pm 0.8 \mu$G -- comparable to that found in dwarf galaxies of a non-tidal origin \citep{chyzy11}\\

Unfortunately, no polarised emission has been found in the present data. One of the possibilities is beam depolarisation, as it was detected neither in the older VLA data at 1.4\,GHz, nor in the NVSS. The angular resolution of the WSRT -- 31\arcsec -- translates into spatial resolution of around 13\,kpc at the distance of the Quintet. This is much more than the expected size of a small TDG.

\subsubsection{NGC\,7320}

It was previously indicated \citep{bnw13b} that it is unclear whether the continuum emission
in the southern part of the group is bound to the Quintet, or it belongs to the interloper
NGC\,7320. Detection of a magnetised outflow that follows the neutral hydrogen tail of the Quintet helps 
to resolve this question at least to the extent that one can be now sure that most of the emission is group-borne.
Unfortunately, the lack of detected polarised emission in this area makes it impossible to employ RM Synthesis for further insights. It is a somewhat surprising result: despite being a dwarf galaxy, the interloper has a well defined spiral structure, and exhibits differential rotation with a sufficiently high maximum rotational velocity \citep{williams02}. Vigorous star-formation, a source of turbulence, was also detected \citep{xu05}. Therefore, all phenomena that are necessary to support the MHD dynamo mechanism \citep{siejkowski11} are present. A possible explanation for the lack of a regular magnetic field might be that the non-rigid rotation crucial for effective dynamo amplification starts outside the star-forming disk, but this is yet to be confirmed.

\subsubsection{NGC\,7317}

The elliptical NGC\,7317 was previously \citep{xu03, osullivan09, bnw13b} believed to be a radio
quiet galaxy. The high sensitivity of the new WSRT data allows us to detect it in our total power map (Fig.~\ref{tp_full}). There are no traces of polarised emission. As early-type galaxies are considered to be lacking mechanisms that could effectively order the magnetic field \citep{beckrev}, this is not an unexpected result. The real character of the bridge that connects NGC\,7317 to the common envelope is uncertain. Whereas this galaxy is believed to be a part of the original triplet of galaxies that formed what is today called the Stephan's Quintet \citep{moles98}, the radio contours are relatively concave on the northern side of the connection area. Keeping in mind the large beam size (comparable to the observed structure) another explanation seems to be more plausible: that this galaxy is detached from the other ones, and the richness of radio emission on its east side originates from an extension of the radio envelope in this direction. 

\subsubsection{NGC\,7331}

The largest and the most luminous member of the Deer Lick Galaxy Group, NGC\,7331, is a well known radio source, and also a host of a regular magnetic field \citep{sings2}. Unfortunately, due to a large angular distance from the phase centre -- approximately 30 arcminutes -- the primary beam response in this area is significantly reduced. Software limitations (\textsc{aips} task \textsc{pbcor}), increased noise in the surrounding area, and overall decrease of the sensitivity (which is frequency-dependent, introducing additional confusion) in the area consistent with the position of NGC\,7331 -- down to a level of around 1\% -- render detailed studies of the morphology of the regular magnetic field in NGC\,7331 impossible from this dataset.

\section{Summary and conclusions}
\label{sec_conc}

Stephan's Quintet is the first galaxy group for which the novel technique of Rotation Measure Synthesis was successfully applied to radio data. This approach has given rise to several interesting discoveries and constraints, which we summarise below:\\

\begin{itemize}
 \item The most important finding is that the medium in the physical vicinity of the Quintet appears to be magnetised, and hosts a large-scale, regular magnetic field. Its extent is at least 60x40\,kpc in the plane of the sky. Owing to the fact that only a product of the strength of regular magnetic field and its depth along the line of sight can be estimated, the exact values of both of these parameters cannot be unambigously established. However, the range of sensible values seems to favour a depth of 10--40\,kpc, and an average field strength of at least 2.5--5\,$\mu$G. Such a strength is of the same order as those found inside spiral galaxies.\\
 \item The linear extent of the newly discovered structure exceeds all previously known regularly magnetised structures found in low-mass galaxy systems.\\
 \item The structure hosting a regular magnetic field is submerged in an even larger region, where
       the magnetic field is dominated by a disorganised component. This region covers all but one member galaxy (although there is an extension towards the latter object), as well as a vast volume
       of neighbouring intergalactic space. On its eastern side, it follows precisely the morphology of the neutral gas tail. On the one hand, all of the structure that emits in the radio continuum seemingly belongs to the Quintet, not to the interloper NGC\,7320. On the other hand, the magnetised matter escapes into the extra-group space through the same outflow, as the neutral one.\\
 \item We considered alternative interpretations involving one or more magnetised screens inside the Milky Way along the line of sight to Stephan's Quintet; however, on the basis of the analysed data, this explanation seems to be less probable -- especially as it is seemingly not concordant with published models of the Galactic magnetic field morphology in the direction of the SQ.\\
 \item We did not detect any polarised emission from the tidal dwarf galaxy SQ--B, which remains in contrast to our previous findings at higher frequencies. The galaxy, is, however, still well visible in the Total Power map from the WSRT. The lack of the polarised detection might be due to beam depolarization generated by the low ($\approx 13$\,kpc) spatial resolution at the distance of the Quintet.\\

\end{itemize}

\begin{acknowledgements}
This work has received support from the Polish National Centre of Sciences (NCN), grant no. UMO-2013/09/N/ST9/02532. The Westerbork Synthesis Radio Telescope is operated by ASTRON (Netherlands Institute for Radio Astronomy) with support from the Netherlands Foundation for Scientific Research (NWO).
\end{acknowledgements}

{}

\end{document}